# Ordered C vacancies in titanium carbides:

# a correlation between crystal structure and the effects on oxidation behavior at elevated temperature

Shaolou Wei[1,†], Lujun Huang[1,*], Yuntong Zhu[2,†], Zhe Shi[3,†] and Lin Geng[1,*]

*1. Key Laboratory of Advanced Structural-Functional Integration Materials & Green Manufacturing Technology, School of Materials Science and Engineering, Harbin Institute of Technology, Harbin, 150001, P.R. China*

*2. School of Materials Science and Engineering, Georgia Institute of Technology, Atlanta, GA 30332, USA*

*3. Department of Materials Science and Engineering, University of Toronto, Toronto, M5S 3E4184, Canada*

†**Present address:** *Department of Materials Science and Engineering, Massachusetts Institute of Technology, Cambridge, MA 02139, USA*

\*Corresponding authors: Lujun Huang (huanglujun@hit.edu.cn); Lin Geng (genglin@hit.edu.cn)

**Abstract**

It has been widely accepted that the introduction of titanium carbides into titanium-based alloys can significantly enhance the oxidation resistance due to their superior physicochemical stability at elevated temperatures. The present study reported for the first time that the ordered C vacancies within titanium carbides could lead to an uncommon phenomenon particularly at the very initial stage of oxidation. The intrinsic micro-to-macro oxidation mechanisms were systematically clarified with the aids of transmission electron microscope and ab-initio molecular dynamics simulation.

Due to their superior specific strength and favorable processability, titanium-based alloys have been considered as an optimal candidate for new-generation spacecraft[1-3]. Apart from the desirable overall mechanical properties, however, their high chemical activity, particularly strong affinity with oxygen at elevated temperatures often lead to early-stage degradation, which consequently hinders them from potential engineering applications[4, 5]. To understand the inherent mechanisms, the motion of oxygen atoms within titanium lattice together with the interactive effects have been investigated from both experimental and computational aspects[6, 7]. Simultaneously, various attempts such as coating, alloying, and introducing secondary particles have been made by experimentalists to improve the oxidation resistance at elevated temperatures for titanium-base alloys[8-10]. Owing to the low cost as well as the convenience in processing, introducing ceramic reinforcements is widely noted as an effective method to fulfill such a purpose[11].

With high melting point, superior thermal stability and outstanding hardness, titanium carbides are promising to further optimize the high-temperature performances for titanium-based alloys [11]. As far as crystal structure is concerned, the most unique feature of titanium carbides is the C vacancies. It was previously reported in the literature that up to 50 % of the C sites can be vacant, and this was widely proved to link with the variation in macroscopic properties such as hardness, elastic modulus and conductivity[12, 13]. Much efforts have also been focused on understanding the physics behind the C vacancies by computational methods[14-16]. Even though previous investigations have revealed the feasibility of improving the oxidation resistance by introducing titanium carbides into titanium alloys[17, 18], little attention was paid to the correlation between C vacancies and macroscopic oxidation behavior. A comprehensive understanding of the oxidation mechanism for titanium carbides reinforced titanium matrix composites remains elusive.

The aim of the present study is to clarify the correlation between the crystal structures of titanium carbides and the oxidation behavior of bulk composites, and to address a micro-to-macro oxidation mechanism with a combination of experiment and simulation. The results of the present work also shed light on the fundamental understanding of performance for metal or ceramic matrix composites as well as $M_{n+1}AX_n$ phases at elevated temperatures.

Fine graphite powders and large spherical Ti-6Al-4V alloy powders were mechanically milled together at a speed of 150 rpm for 5 hrs under protective high-purity argon gas. To fabricate bulk Ti-6Al-4V/TiC composites, the milled mixtures were subsequently reaction hot pressed sintered at 1473 K for 1 h under a pressure of 25 MPa with vacuum rate below $10^{-3}$ Pa. In-situ chemical reaction between Ti and C took place spontaneously during such process, leading to the formation of TiC, of which the detailed mechanisms were clarified elsewhere[11]. The same processing technique was also employed to achieve the monolithic Ti-6Al-4V alloy. The specimens of dimensions 10mm×10mm×3mm for oxidation experiments were wire cut from both alloy and composites. They were ground on a series of SiC abrasive papers up to 2000# and then ultrasonically cleaned for 5 min in acetone bath. The oxidation behavior was investigated under the conditions of isothermal static air at 873 K for 20 hrs inside a heat-treatment furnace, during which the mass change was monitored by a PTX-FA210 electrical balance with precision of 0.1 mg. Mass change values of three specimens were averaged each time so as to avoid experimental occasstionality. The oxidized specimens were analyzed by a Zeiss SUPRA 55 SAPPHIR scanning electron microscope (SEM) equipped with an INCA 300 energy dispersive spectrometer (EDS) and a Dimension Fastscan atomic force microscope (AFM). An FEI Talos 200 transmission electron microscope (TEM) was employed to accomplish crystallographic characterizations. All the TEM specimens were prepared through ion beam milling methods.

The DFT simulations were performed by the Vienna Ab Initio Simulation Package[19] and the Perdew-Burke-Ernzerhof generalized-gradient approximation functional[20] was used for all calculations. Supercells each with 32 Ti atoms, were built up for TiC, $TiC_{0.5}$ and α-Ti and were subject to structural relaxation at first. A Monkhorst−Pack k-point mesh of 9 ×9 ×9 was applied with an energy cutoff of 800 eV and a maximum force tolerance on each atom of 0.02 eV/Å. When performing ab initio molecular dynamics (MD) simulation at finite temperature, the same k-point mesh and an energy criterion of $1 \times 10^{-5}$ eV were adopted to ensure high computational accuracy. All the MD simulations were conducted within an NVT ensemble at the experimental temperature of 873 K using a Nose-Hoover thermostat[21]. The simulations were performed with a time step of 1 fs until the total energy fluctuation was kept within 0.02 eV for at least 2 ps.

On the basis of classical thermodynamics, the Gibbs free energy change ($\Delta G_r$) for the oxidation reactions of TiC (TiC+$O_2$→rutile+$CO_2$) and α-Ti (Ti+$O_2$→rutile) were calculated respectively. As depicted in Fig. 1 (a), it is evident that the ceramic phase TiC exhibits superior oxidation resistance to Ti, which coincides with the results reported by Qin et al[22]. The EDS analyses of oxygen content on titanium carbides as well as the matrix α-Ti, however, indicate the existence of both normal and abnormal behavior of the carbides at the very initial stage of oxidation, as in Fig. 1(b). The oxygen content on some carbides (53.54 wt.%) is lower than that on the matrix (61.21 wt.%), agreeing well with the calculations. Nonetheless, the opposite case is also observed: the oxygen content on the carbides (46.99 wt.%) is significantly higher than that on the matrix (29.69 wt.%), which goes directly against both the thermodynamic predictions and the conventional understandings. No apparent morphological distinctions can be found once comparing the representative SEM images for those two kinds of titanium carbides, as in Fig. 1 (c) and (d).

Fig. 2 demonstrates the TEM studies of titanium carbides particles within the composites. Selected area electron diffraction (SAED) patterns clearly reveal the intrinsic differences in crystal structure: Fig. 2 (a1) displays typical perfect face-centered cubic (FCC) diffraction patterns corresponded to [100] zone axis, which is similar to the investigations reported in the literature[23]. While in Fig. 2 (b1), <1/2, 1/2, 1/2> type superlattic spots can be clearly observed, and this directly supports the existence of ordered microdomains. Compared the SAED patterns in Fig. 2(b1) to the study accomplished by Bursik et al[24], it can be concluded that the corresponding C/Ti ratio is 0.5 ($TiC_{0.5}$ in chemical formula). It should also be noted that no other types of superlattice spots are detected in our present work. Both TiC and $TiC_{0.5}$ bond well with the adjacent α-Ti matrix and no traits of interfacial reactions are found, which is confirmed by the TEM bright field images together with EDS elemental distribution analyses presented in the rest of Fig. 2.

It is now clear from the TEM results that two kinds of carbides exist in the bulk composites: conventional TiC and $TiC_{0.5}$ (with ordered C vacancies). On the basis of the evident distinctions in crystal structure, the following micro-to-macro oxidation mechanisms are proposed. Microscopically, at the very initial stage of oxidation, the thermally activated oxygen atoms tend to preferentially occupy the ordered C vacancies within $TiC_{0.5}$, resulting in the abnormally high oxygen content on it, which is against the thermodynamic predictions; macroscopically, there exists a critical point in the mass gain kinetics, before which the easily oxidized $TiC_{0.5}$ will lead to higher mass-gain rate for the composites; after which, however, the overall physicochemical properties of titanium carbides take charge of the oxidation kinetics and consequently lead to the decrease in mass gain rate.

To prove the microscopic validity, ab-initio MD simulation was employed to capture the energy variation after O atoms occupy the interstices in TiC (tetrahedral site) and $TiC_{0.5}$ (octahedral site) and α-Ti (octahedral site). The energy of TiC and $TiC_{0.5}$ is lowered by 1.14 eV (0.19%) and 9.62 eV (2.27%) after introducing one O atom to the lattice, comparing to an energy decrease of 8.65 eV (1.76%) for α-Ti, which is depicted in Fig.3 (a). The relatively larger energy decrease of $TiC_{0.5}$ with one interstitial O indicates the energetically favored occupation of the ordered C vacancies, which therefore confirms the microscopic explanation of abnormal high oxygen content discussed above.

To prove the macroscopic validity, the isothermal mass-gain kinetics curves of the monolithic Ti-6Al-4V alloy and Ti-6Al-4V/TiC composites were measured. As presented in Fig. 3(b), the kinetic characteristics for these two materials both yield the parabolic principle, indicating the formation of protective oxide scales. At the very early stage (especially before 5 hrs), however, the Ti-6Al-4V/TiC composites possess higher mass-gain rate than the monolithic Ti-6Al-4V alloy, this is because the O atoms preferentially diffuse into the ordered C vacancies traps within $TiC_{0.5}$ so as to minimize the total energy, which was proved quantitatively by the ab-initio MD simulation.

With oxidation time extends, the mass-gain rate for Ti-6Al-4V/TiC composites decreases significantly when compared to that of Ti-6Al-4V alloy, leading to final-stage mass-gain values of 1.57 mg/cm$^2$ and 1.99 mg/cm$^2$, respectively. Such enhanced long-term isothermal oxidation resistance of Ti-6Al-4V/TiC is attributed to the formation of denser and finer oxide scales, which is proved by the AFM analyses (inset images) of both alloy and composites after 20 hrs oxidation. In addition, TiC phase plays an important role in enhanced oxidation resistance due to lower energy decrease after the inset of O atoms compared with that of the matrix titanium alloy. It is worthwhile noting that the critical point in the present study is around 4 hrs, before which the unique crystal structure of TiC$_{0.5}$ control the kinetic characteristic, while with increasing oxidation time, the overall physicochemical properties of titanium carbides together with finer and denser oxide scales become the dominant factor.

In conclusion, the oxidation behavior of Ti-6Al-4V/TiC composites not only relies on the physicochemical properties but also the intrinsic crystal structure of titanium carbides: O atoms tend to preferentially occupy the ordered C vacancies within TiC$_{0.5}$ resulting in the abnormally high O content. Simultaneously, the easily oxidized TiC$_{0.5}$ also leads to a faster mass-gain trend for Ti-6Al-4V/TiC composites at the very initial stage. With extending oxidation time, it is the overall stability of titanium carbides and the finer oxide scales on Ti-6Al-4V/TiC composites that dominates the oxidation process, which in turn accounts for the long-term superior isothermal oxidation resistance.

**Acknowledgements**

This work was financially supported by National Key R&D Program of China (No. 2017YFB0703100) and National Natural Science Foundation of China under the grant numbers of Nos. 51671068 and 51471063. The authors are sincerely grateful to Prof. Jian Chang, Prof. Wei Zhai and Prof. Guohua Fan for their enthusiastic help with the experiments. Shaolou Wei would like to thank Dr. Mohadeseh Taheri Mousavi for critical suggestions on ab-initio MD simulation.

**Figures and captions:**

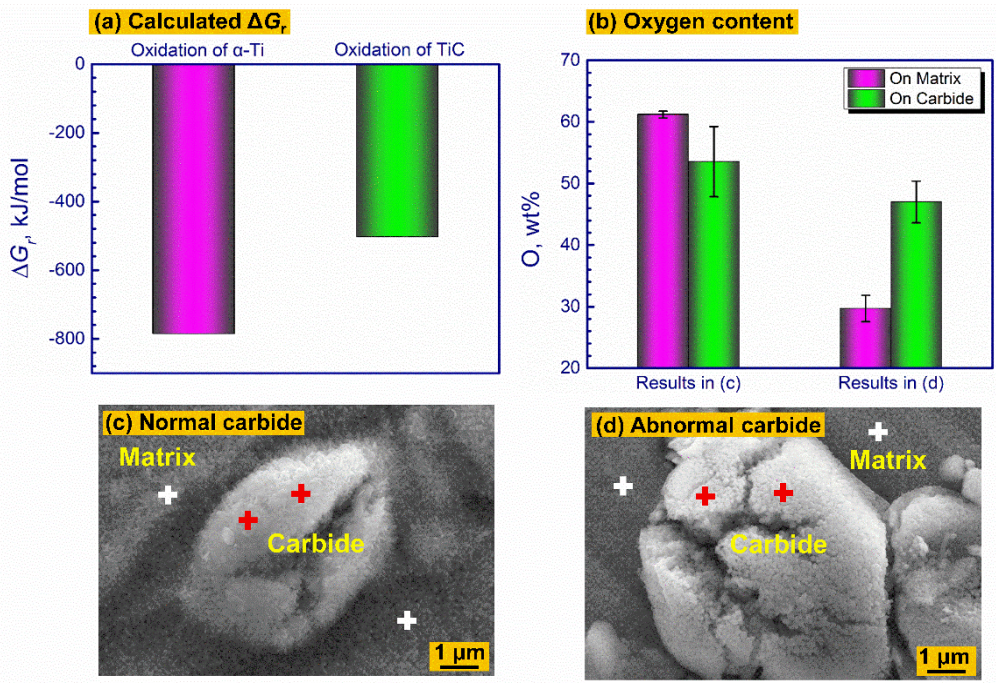

Fig.1 (a) Calculated Gibbs free energy change for the oxidation of α-Ti and TiC; (b) measured oxygen content on titanium alloy matrix and titanium carbides; (c) and (d) SEM images of the normal and abnormal titanium carbides after oxidation for 5 hrs at 873 K (tested region marked with crosses)

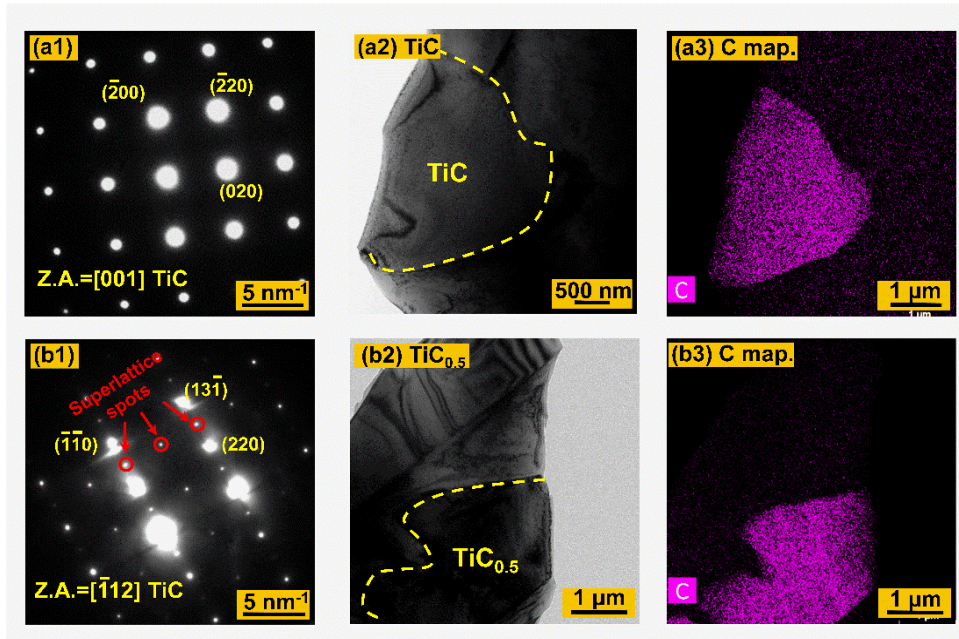

Fig. 2 Typical TEM results for two kinds of titanium carbides within the composites: (a1) and (b1) SAED patterns; (a2) and (b2) bright field images; (a3) and (b3) C mapping

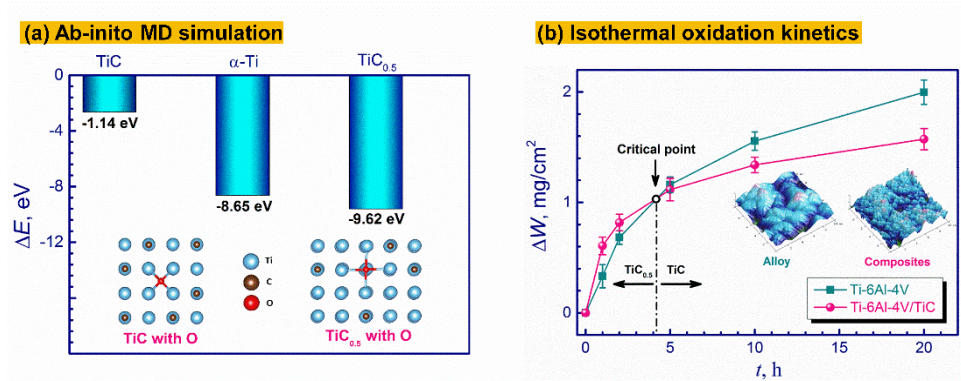

Fig. 3 (a) Ab-initio MD simulation results for energy decrease after introducing O atom; (b) isothermal oxidation kinetic curves

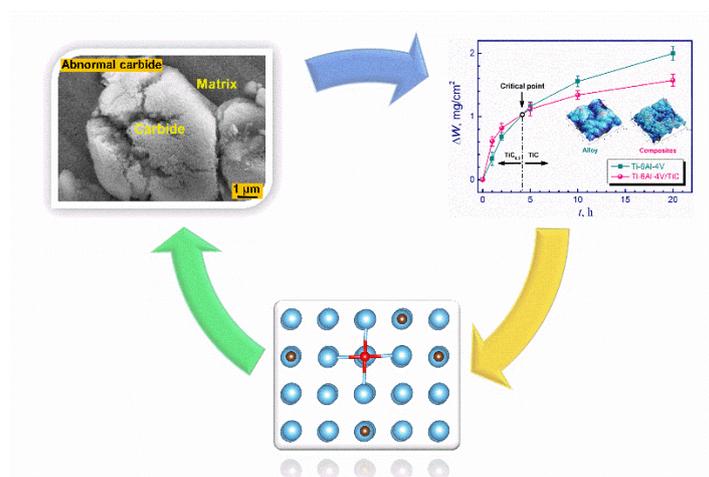

Graphical abstract